\documentclass[twocolumn,prl,superscriptaddress,aps,amsmath,floatfix,longbibliography]{revtex4-2}

\usepackage[pdfborder={0 0 0},colorlinks=true,linkcolor=blue,urlcolor=blue,citecolor=blue]{hyperref}

\usepackage{graphics}
\usepackage{graphicx}
\usepackage{epstopdf}
\usepackage{dcolumn}
\usepackage{bm}
\usepackage{longtable}
\usepackage{epsfig}
\usepackage{times}
\usepackage{url}

\usepackage{color}

\usepackage{color,soul}

\begin{document}
\title{Generation and Coherent Control of Dark-State Spatial Modes}

\author{Huai-Che \surname{Shia}}
\affiliation{Department of Physics, National Central University, Taoyuan City 32001, Taiwan}
\author{Siang-Wei \surname{Shao}}
\affiliation{Department of Physics, National Central University, Taoyuan City 32001, Taiwan}
\author{Wu-Cheng \surname{Chiang}}
\affiliation{Department of Physics, National Central University, Taoyuan City 32001, Taiwan}
\author{Yu-Hung \surname{Kuan}}
\affiliation{Department of Physics, National Central University, Taoyuan City 32001, Taiwan}
\author{I-Kang \surname{Liu}}
\affiliation{School of Mathematics, Statistics and Physics, Newcastle University, Newcastle upon Tyne, NE1 7RU, United Kingdom}
\author{Teodora \surname{Kirova}}
\affiliation{Institute of Atomic Physics and Spectroscopy, Faculty of Science and Technology, University of Latvia, LV-1004 Riga, Latvia}
\author{Gediminas \surname{Juzeli\=unas}}
\affiliation{Institute of Theoretical Physics and Astronomy, Department of Physics, Vilnius University, Saulėtekio 3, Vilnius, LT-10257, Lithuania}
\author{Yu-Ju \surname{Lin}}
\affiliation{Institute of Atomic and Molecular Sciences, Academia Sinica, Taipei 10617, Taiwan}
\author{Wen-Te \surname{Liao}}
\email{wente.liao@g.ncu.edu.tw}
\affiliation{Department of Physics, National Central University, Taoyuan City 32001, Taiwan}
\affiliation{Physics Division, National Center for Theoretical Sciences, Taipei 10617, Taiwan}
\affiliation{Quantum Technology Center, National Central University, Taoyuan City 32001, Taiwan}
\date{\today}
\begin{abstract}
The generation and dynamic control of the spatial mode of the dark-state polarization using electromagnetically induced transparency are theoretically investigated.
We demonstrate that a combination of synthetic scalar  and vector potentials can be employed  to engineer discrete spatial modes of the dark state polariton, enabling quantum interference among these modes.
We verify this concept by showing the Rabi oscillation between two spatial modes and stimulated Raman adiabatic passage among $\Lambda$-type three modes.  Our approach allows for the reallocation of stored photonic data from one location to another, presenting potential applications such as photonic memory optimization and retrieved light modulation.
\end{abstract}

\maketitle
%
%

Dark-state polaritons, electrically neutral quasiparticles arising from light-matter excitations, are central to Electromagnetically Induced Transparency (EIT), where a strong coupling field renders a medium transparent by altering its properties \cite{Harris1997, Harris1990, fleischhauer2000}. EIT enables slow light \cite{Liu2001, Phillips2001, Kocharovskaya2001}, facilitating applications in light storage \cite{Phillips2001, Fleischhauer2005}. Stationary-light polaritons, a specialized form, occur in double-atomic systems interacting with counter-propagating weak fields and control lasers, behaving as two-component Dirac particles with an effective mass controlled by the field \cite{bajcsy2003, Zimmer2008, Lin2009, Otterbach2010, unanyan2010}. This opens possibilities in nonlinear optics, quantum information, and particle effects in effective magnetic fields. Concurrently, stimulated Raman adiabatic passage (STIRAP), introduced by K. Bergman’s team as an efficient technique for selective quantum state population transfer with minimal loss \cite{Gaubatz1990}, has applications in atomic and molecular optics, ultracold physics, and quantum information \cite{vitanov2017}. 

Combining EIT and STIRAP, this work develops a method for generating and dynamically controlling dark-state spatial modes (DSMs) using synthetic scalar and vector potentials, offering prospects for fundamental quantum studies and practical applications like memory optimization and modulation.
Our idea is inspired by transferring the coupled atom-light equations, i.e., optical Bloch equation (OBE) \cite{fleischhauer2000, bajcsy2003, Lin2009, Otterbach2010, Kuan2023} presented below, into a  Schr\"odinger-like equation of the dark-state  polarization  \cite{Otterbach2010, Kuan2023}.
Through this transformation, we can find out the relationship between single particle quantum system and our EIT system. 
We further validate this transformation by numerically solving the full OBE and compare the results with the predictions from the Schr\"odinger-like equation. Specifically, we demonstrate that a combination of synthetic scalar and vector potentials can be used to engineer discrete spatial modes of the dark-state polarization, allowing for the reconstruction of arbitrary quantum-state series and the control of quantum interference among these modes.

\begin{figure}[b]
\includegraphics[width=0.49\textwidth]{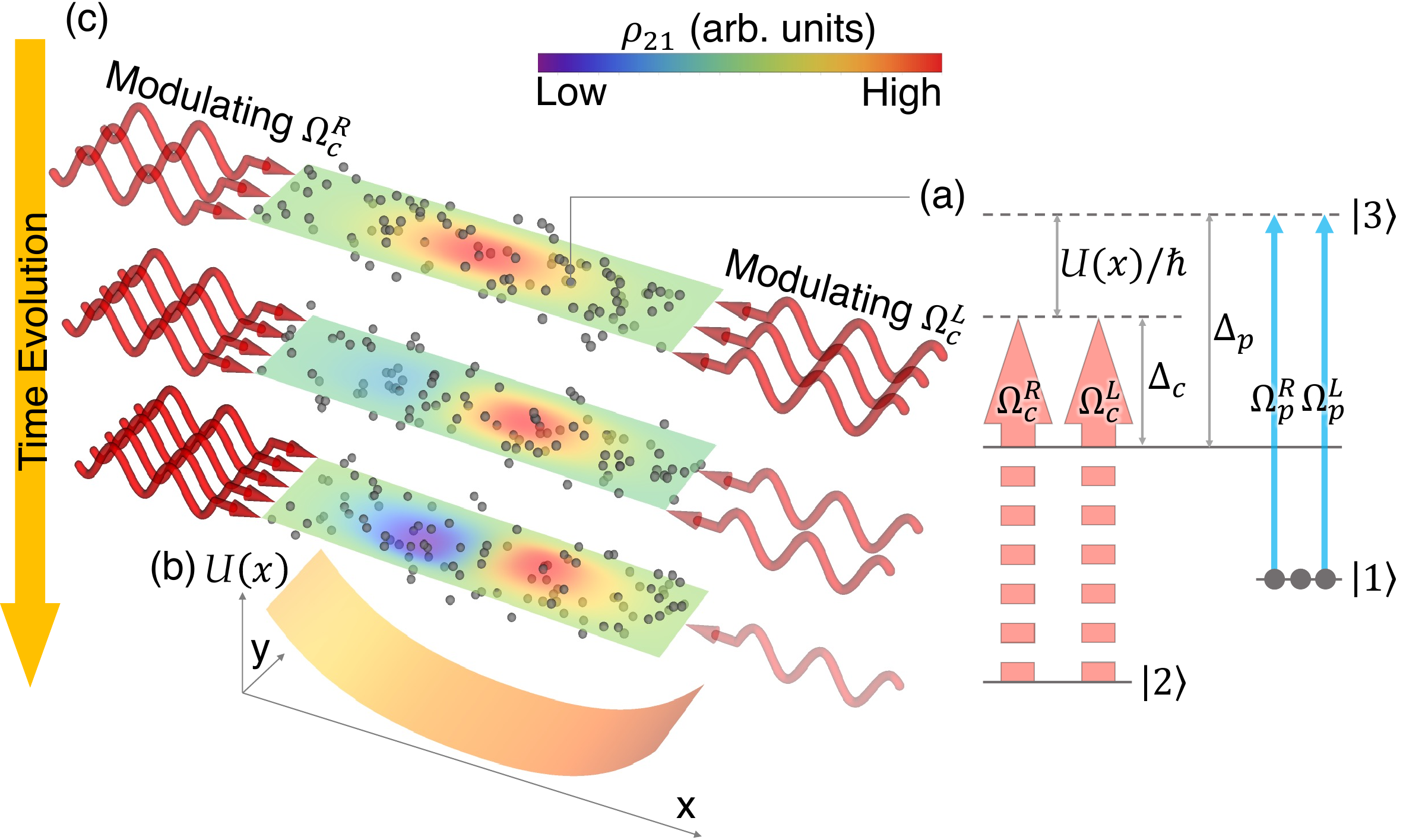}
\caption{\label{fig1}
(a) Three-level-$\Lambda$-type EIT system. Blue-upward arrows depict the probe fields. The red-thick-dashed-upward arrows denote the periodically modulating $\Omega_c^R$ and $\Omega_c^L$. 
The two-photon detuning $\Delta_p-\Delta_c$ forms the synthetic scalar potential energy $U(\mathbf{x})$ as depicted in (b).
(c) The sketch of the Schr\"odinger equation simulator using EIT. Gray dots represent atoms, and the red-sinusoidal arrows illustrate two counter-propagating control fields. The density and the opacity of arrows reflect the control field strength.
The coloured density plot depicts  the dark-state polarization $\rho_{21}$.
}
\end{figure}
Figure~\ref{fig1} illustrates our quasi-one-dimensional counter-propagating three-level-$\Lambda$-type EIT system, which is described by the  OBE \cite{fleischhauer2000, bajcsy2003, Lin2009, Otterbach2010, Hsiao2018, Kuan2023}
\begin{eqnarray}
\partial_t \rho_{21} &=& 
	\left[ i \left( \Delta_c - \Delta_p\right) -\gamma \right]   \rho_{21} 
	+  \frac{i}{2} \left( \Omega_c^{R*} \rho_{31}^R + \Omega_c^{L*} \rho_{31}^L \right)   , \label{OBE21}\\
\partial_t \rho_{31}^R &=& \frac{i}{2} \Omega_p^R + \frac{i}{2} \Omega_c^R \rho_{21} -\left(  \frac{\Gamma}{2} +i\Delta_p \right) \rho_{31}^R, \label{OBE31R}\\
\partial_t \rho_{31}^L &=& \frac{i}{2} \Omega_p^L + \frac{i}{2} \Omega_c^L \rho_{21} -\left(  \frac{\Gamma}{2} +i\Delta_p \right) \rho_{31}^L, \label{OBE31L}\\
\frac{1}{c} \partial_t \Omega_p^R &+& \partial_x \Omega_p^R = i \eta \rho_{31}^R, \\
\frac{1}{c} \partial_t \Omega_p^L &-& \partial_x \Omega_p^L = i \eta \rho_{31}^L.\label{OBEprobeL}
\end{eqnarray}
$\rho_{ij}$ is the density matrix element of the three-level atom  in Fig.~\ref{fig1}(a).
A pair of counter-propagating control (probe) fields drives the $\vert 2 \rangle \rightarrow \vert 3 \rangle$ $\left( \vert 1 \rangle \rightarrow \vert 3 \rangle \right)$ atomic  transition with Rabi frequency $\Omega_c^{R}$ and $\Omega_c^{L}$ ($\Omega_p^{R}$ and $\Omega_p^{L}$).
Superscript $R$ and $L$ denote the right- and left-going fields.
$\Delta_c$ ($\Delta_p$) is the one-photon detuning of  the control (probe) field.
$\eta=\frac{\Gamma\xi}{2 L}$  is the light-matter coupling constant. $\Gamma$ is the spontaneous decay rate of the  excited state $\vert 3\rangle$, and $\gamma$ is the the dephasing rate between the ground states.   
$\xi$  and $L$  are  the optical depth and  the medium length, respectively. 
Our system is quasi-one-dimensional, because all control fields and detunings are uniform in $\mathbf{y}$ direction but vary  along only $\mathbf{x}$ direction.

Transferring Eq.~(\ref{OBE21}-\ref{OBEprobeL}) into 
a  Schr\"odinger-like equation of the dark-state  polarization $\rho_{21}$ \cite{Otterbach2010, Kuan2023}
\begin{equation}
i\hbar\partial_t \rho_{21} = \frac{\left( \frac{\hbar}{i}  \partial_x + A_x \right) ^2}{2\mathsf{m}} \rho_{21}
+U\rho_{21} + i \left( \frac{\Gamma}{2\Delta_p}\right)\frac{\hbar^2}{2\mathsf{m}} \partial_x^2 \rho_{21}.
\label{schrodinger}
\end{equation}
Here, $\rho_{21}$ serves as the wave function,
and $\hbar$ is the reduced Planck constant.
The x component of the synthetic vector potential $A_x$ and scalar potential energy $U$, are given by \cite{Kuan2023}. 
\begin{eqnarray}
A_x &=& \mathsf{m} V_g ,\label{eq2} \\
U &=& \hbar \left( \Delta_p-\Delta_c -i\gamma\right)  -  \frac{1}{2 \mathsf{m}}\vert A_x \vert^2. \label{eq3}
\end{eqnarray}
The product of  the effective mass $\mathsf{m}=\frac{\hbar\eta^2}{2\Delta_p\Omega_c^2}$ \cite{unanyan2010, Kuan2023}, where $\Omega_c^2=\vert \Omega_c^{R}\vert^2 + \vert \Omega_c^{L}\vert^2$,  and the EIT group velocity  $V_g = \frac{1 }{2\eta}\left( \vert \Omega_c^{R}\vert^2 - \vert \Omega_c^{L}\vert^2\right)$ represents the vector potential $A_x$ for a unit charge. 
To maintain a constant effective mass ($\mathsf{m}$), we keep $\Omega_c$ invariant. 
Under these conditions, the $\vert  A_x \vert^2$ term in the scalar potential $U(\mathbf{x})$ is typically negligible in quantum optics. Consequently, the two-photon detuning $\Delta_p-\Delta_c$ becomes the dominant contributor to $U(\mathbf{x})$ as illustrated in Fig.~\ref{fig1}(b).
The rightmost term in \eqref{schrodinger} represents diffusion, which can be mitigated by increasing the size of the probe field.

Equation (\ref{schrodinger}) shows that spatial eigenmodes of $\rho_{21}$ can be generated for different $U(\mathbf{x})$ and coherently coupled using the synthetic time-varying electric field $E_x = -\partial_t A_x$.
The preparation of any initial wave function $\rho_{21}\left( \mathbf{x}, 0\right) $ can be implemented  by invoking EIT light storage \cite{Liu2001,Phillips2001,Kocharovskaya2001}. 
First, the control laser is turned off to store a probe field with the profile of the $n$th eigenmode $\Psi_n\left(  \mathbf{x}\right) $ for a given $U(\mathbf{x})$. 
Then, by turning on two counter-propagating control fields $\Omega_c^R = \Omega_c^L = \frac{\Omega_c}{\sqrt{2}}$ with $\Delta_p - \Delta_c =  \frac{U(\mathbf{x})}{\hbar}$, the stationary light is retrieved, resulting in the desired DSMs $\rho_{21}\left( \mathbf{x}, 0\right) = \Psi_n\left(  \mathbf{x}\right) $ \cite{bajcsy2003, Lin2009}.
Finally, the coherent transfer between DSMs $\Psi_n$ and $\Psi_m$ can be achieved by dynamically modulating the intensity of the two control fields, as illustrated in Fig.~\ref{fig1}(c).
We will demonstrate the Rabi oscillation between two DSMs, and STIRAP among three DSMs \cite{vitanov2017, bergmann2019, liao2011}
of the Fluxonium potential $ U(\mathbf{x}) = \hbar C_1 [\mathbf{x}^2 + \alpha^{2} \cos\left( \frac {2 \pi  \mathbf{x}}{\lambda_{f}}+ \phi \right)]$ \cite{manucharyan2009}.
By adjusting parameters $\left( C_1,\alpha,\lambda_{f},\phi \right) $, the DSMs can be precisely engineered to suit specific experimental or theoretical requirements.
Our numerical simulations use the following common parameters \cite{Hsiao2018}:
$\Gamma = 2\pi \times 5.2$ MHz,\
$\Delta_p = 4.6 \Gamma$, 
$\gamma = 10^{-3} \Gamma$, and
$\xi = 800$.

\begin{figure}[t]
\includegraphics[width=0.5\textwidth]{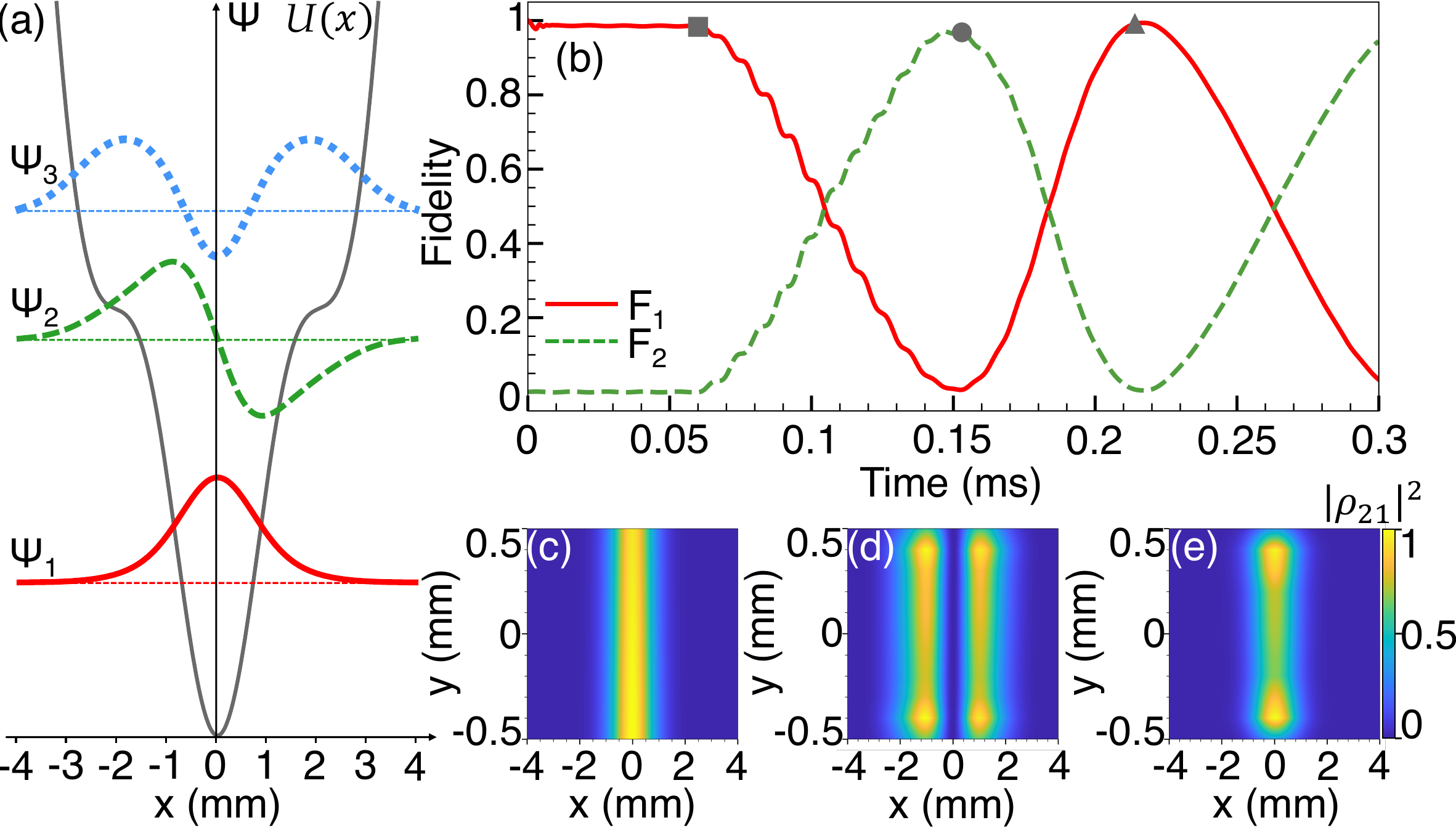}
\caption{\label{fig2}
Illustration of the Rabi oscillation scheme. 
(a) Gray line represents the synthetic scalar potential  $U(\mathbf{x})$.  Thick curves denote the $n$th eigenmode $\Psi_n$ and thin lines represent the corresponding eigenenergy  $\hbar\omega_n$.
(b) Time evolution of fidelity $F_n$ of the  $n$th DSM. Red-solid (green-dashed) line denotes $F_1$ ($F_2$). The snapshot of  $\vert \rho_{21}\vert^2$ at (c) $t=0.06$ms, (d) $t = 0.15$ms, and (e) $t=0.21$ms, as respectively indicated by the gray square, dot, and  triangle in (b).
}
\end{figure}
We now demonstrate the Rabi oscillation with $\left( C_1,\alpha,\lambda_f,\phi\right) = ( 3.06\times 10^{-3} \frac{\Gamma}{\text{mm}^2}, \sqrt{1.8} \text{ mm}, 2.89\text{ mm},$
$3.14) $,
$L = 8$ mm,
and $\Omega_c = 1.5 \Gamma$.
%
%
The $n$th eigen-wavefunction $\Psi_n\left( \mathbf{x}\right) $ and its eigen-angular frequency $\omega_n$ are obtained by solving the eigenvalue problem of \eqref{schrodinger} without $A_x$ \cite{griffiths2018}.
Fig.~\ref{fig2}(a) illustrates the potential $U(\mathbf{x})$ (gray-solid line)  and its first three DSMs: $\Psi_1$ (red-solid line), $\Psi_2$ (green-dashed line), and $\Psi_3$ (blue-dotted line).
The eigen-angular frequencies are
$\omega_1 = 232$ rad$\cdot$kHz, 
$\omega_2 = 601$ rad$\cdot$kHz, and 
$\omega_3 = 796$ rad$\cdot$kHz.
To coherently couple these DSMs, we follow the guidance of  \eqref{schrodinger}. Specifically, we introduce time-dependent control fields
$\Omega_c^R\left( t\right) = \frac{\Omega_c}{\sqrt{2}} \sqrt{1 + \beta \sin\left( \nu t \right)}$ and $\Omega_c^L\left( t\right) = \frac{\Omega_c}{\sqrt{2}} \sqrt{1 - \beta \sin\left( \nu t \right)}$ 
with $\vert\beta\vert<1$
to generate 
$A_x\left( t\right) = \beta \frac{\hbar \eta}{4 \Delta_p} \sin\left( \nu t \right)$, while ensuring a constant effective mass.
%
One can match $\nu \approx \omega_m - \omega_n$  to drive   $\Psi_{m} \rightarrow \Psi_{n}$ transition and calculate the effective driving Rabi frequency $\Omega_{nm} = \frac{\beta\Omega_c^2}{2 i \eta} \int_{-\infty}^\infty \Psi_{n}^\ast   \partial_x  \Psi_{m} d\mathbf{x}$ for our quasi-one-dimensional system  \cite{scully1997}.
To explore the dynamics, we numerically solve the full  Eqs.~(\ref{OBE21}-\ref{OBEprobeL}).
As shown in Fig.~\ref{fig2}(b-e), the periodic modulation of the control fields dynamically alters the EIT group velocity and thus effectively shakes the dark-state polarization 
$\rho_{21}$ back and forth within the trapping potential  $U(\mathbf{x})$. 
In Fig.~\ref{fig2}(b), we analyse the mode fidelity at $\mathbf{y} =0$:
\begin{equation}
F_n(t) = \frac{ \left| \int_{-\infty}^{\infty}  \Psi_n^\ast \left( \mathbf{x} \right) \rho_{21}(\mathbf{x},t)  \, d\mathbf{x} \right|^2}{ \left( \int_{-\infty}^{\infty} \left| \Psi_n \left( \mathbf{x} \right) \right|^2 \, d\mathbf{x} \right) \left( \int_{-\infty}^{\infty} \left| \rho_{21}(\mathbf{x},t) \right|^2 \, d\mathbf{x} \right)}.     \nonumber
\end{equation}
The initial condition $\rho_{21}\left( \mathbf{x}, 0 \right)  = \Psi_1 \left( \mathbf{x} \right)$ is prepared   through EIT light storage and retrieval. 
Here we use $\beta= 0.07$ and $\nu = 376.4$ rad$\cdot$kHz for the modulation of the  of the control laser intensity, and the modulation begins at $t = 0.06$ms.
As predicted by \eqref{schrodinger}, the Rabi oscillation between $\Psi_1 $ and $ \Psi_2$ occurs. 
Fidelity $F_1$ drops from 1 to 0 in approximately $0.09$ ms, while $F_2$ increases from 0 to 1 over the same interval. 
The period of the Rabi cycle aligns with the theoretical prediction of $2\pi / \Omega_{21} \approx 0.15$ms.
To further illustrate the strong coupling effect, Fig.~\ref{fig2}(c-e) present the snapshot of $\vert \rho_{21} \vert^2$ at $t = 0.06$ ms, $ 0.15$ ms, and $0.21$ ms, respectively.
The alternating patterns between $\vert \Psi_1 \vert^2$ and $\vert \Psi_2 \vert^2$ confirm the strong coherent coupling between the two modes.
Notably, at $t=0.15$ms when $F_2$ also reaches its maximum, $\vert\rho_{21}\left( \mathbf{x}, t \right)\vert^2$ becomes $ \vert\Psi_2 \left( \mathbf{x} \right)\vert^2$,  characterised by the double-hump structure of the first excited DSM.
%

\begin{figure}[t] 
\includegraphics[width=0.5\textwidth]{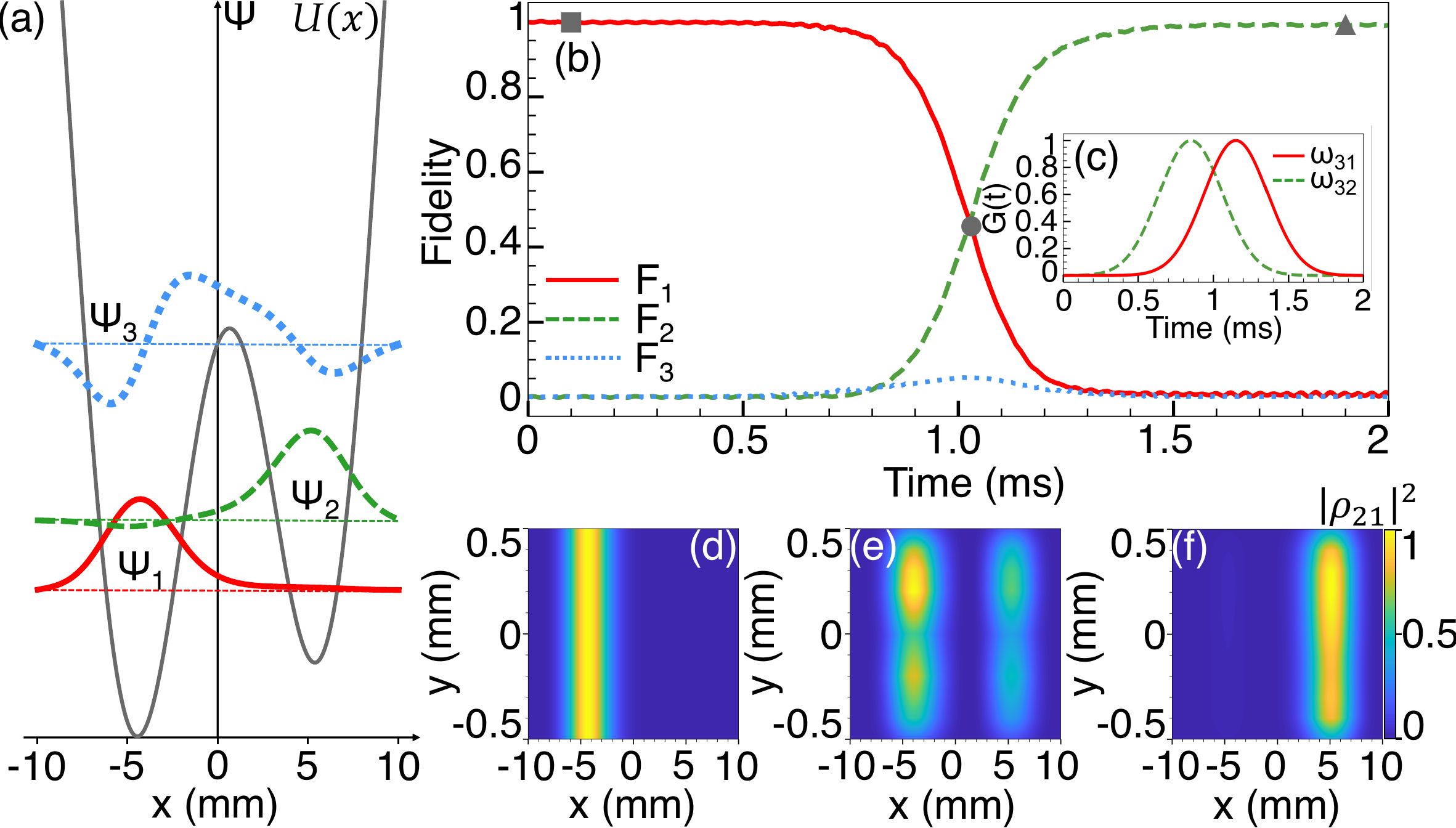} 
\caption{\label{fig3} 
STIRAP  with non-degenerate ground modes. 
(a) The gray line shows an asymmetric $U(\mathbf{x})$. The thick red-solid, green-dashed, and blue-dotted lines depict the DSMs $\Psi_1$, $\Psi_2$, and $\Psi_3$, respectively, and the thin lines denote their eigenenergy.  
(b) Evolution of $F_1$ (red-solid line), $F_2$ (green-dashed line), and $F_3$ (blue-dotted line). 
(c) Two Gaussian envelopes $G\left(t \right)$ of each control field. 
Red-solid (green-dashed)  line drives $\Psi_1 \rightarrow \Psi_3$ ($\Psi_2 \rightarrow \Psi_3$) transition. 
The snapshot of  $\vert \rho_{21}\vert^2$ at (d) $t=0.1$ms,  (e) $t = 1.0$ms, and (f) $t=1.9$ms, as indicated by the gray square, dot, and  triangle, respectively, in (b). 
}
\end{figure}

We now discuss the STIRAP modelling with $\left(C_1,\alpha,\lambda_f,\phi)=(6.12 \times 10^{-4} \frac{\Gamma}{\text{mm}^2}, \sqrt{40} \text{ mm}, 11.56\text{ mm}, 6 \right)$. To maintain diffusion at the same order, we set $L=2$ cm, and $\Omega_c=2 \Gamma$. Fig.~\ref{fig3}(a) illustrates our design of $U(\mathbf{x})$ (gray-solid line)  along with its first three DSMs with their eigen-angular frequencies 
$\omega_1 = 407$ rad$\cdot$kHz, 
$\omega_2 = 601$ rad$\cdot$kHz, and 
$\omega_3 = 1091$ rad$\cdot$kHz.
Unlike the single-well potential in Fig.~\ref{fig2}(a), we utilize a double-well potential to generate a typical $\Lambda$-type system suitable for STIRAP.
The potential is optimized under the condition $\vert\Omega_{21}\vert\ll\vert\Omega_{31}\vert$ and $\vert\Omega_{21}\vert\ll\vert\Omega_{32}\vert$, yielding the effective Rabi frequency $\vert\Omega_{21}\vert = 45$ rad$\cdot$kHz, $\vert\Omega_{31}\vert = 475$ rad$\cdot$kHz, and $\vert\Omega_{32}\vert = 282$ rad$\cdot$kHz.
The central potential barrier  separates  $\Psi_1$ (red-solid line) and $ \Psi_2$ (green-dashed line), making the direct transition $\Psi_{1} \rightarrow \Psi_{2}$ dipole-forbidden.
Besides, the mode $\Psi_3$ (blue-dotted line) partially overlaps with both $\Psi_1$ on the left and $\Psi_2$ on the right, enabling dipole-allowed transitions $\Psi_1 \rightarrow \Psi_3$ and $\Psi_2 \rightarrow \Psi_3$. 
%
%
\begin{figure}[t]
\includegraphics[width=0.5\textwidth]{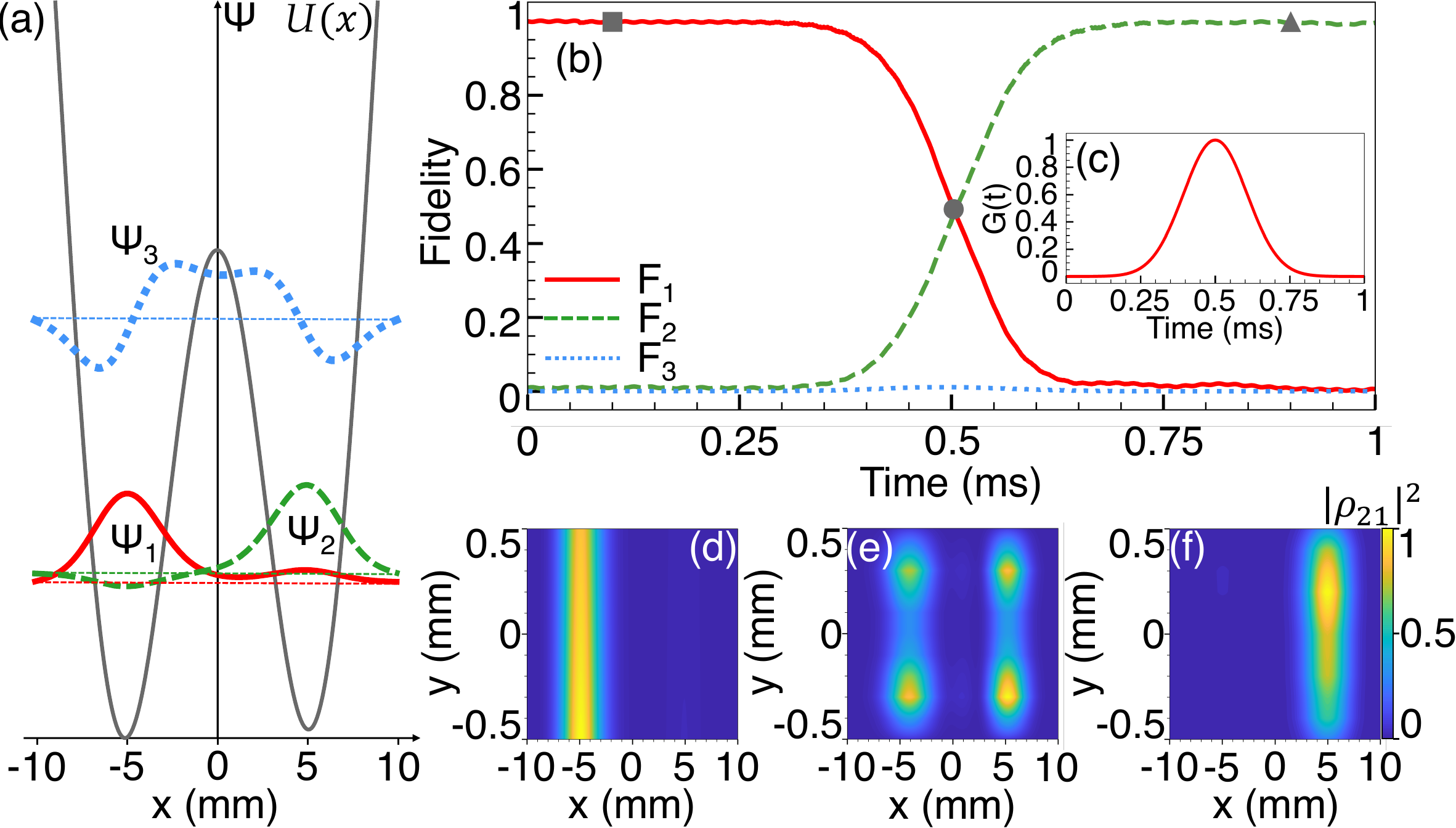}
\caption{\label{fig4}
STIRAP  with degenerate ground modes. 
(a) The gray line depicts a symmetric $U(\mathbf{x})$, and the thick colored lines show its DSMs. $\Psi_1$ and $\Psi_2$ are degenerate. As for slender lines denotes their DSM levels $\omega_1$, $\omega_2$, and $\omega_3$.
(b) Red-solid, green-dashed, and blue-dotted lines shows the time evolution of $F_1$, $F_2$, and $F_3$, respectively.
(c) The Gaussian modulation envelope $G\left(t \right)$ of each control field. 
Red-solid  line drives  both $\Psi_1 \rightarrow \Psi_3$ and $\Psi_2 \rightarrow \Psi_3$ transition.
The snapshot of  $\vert \rho_{21}\vert^2$ at (d) $t=0.1$ms, (e) $t = 0.5$ms, and (f) $t=0.9$ms, as specified by the gray square, dot, and  triangle, respectively, in (b).
}
\end{figure}
To drive a Raman transition mediated by the dipole-allowed routes, we modulate the control lasers using two adjacent Gaussian profiles:
\begin{eqnarray}
\Omega_c^R\left( t\right)  &=& \frac{\Omega_c}{\sqrt{2}} \sqrt{1 + \sum_{j=1}^2\beta_j \sin(\nu_j t)\exp\left[-\left(\frac{t - \epsilon_j}{\tau}\right)^2\right] }, \nonumber\\
\Omega_c^L\left( t\right) &=& \frac{\Omega_c}{\sqrt{2}} \sqrt{1 - \sum_{j=1}^2\beta_j \sin(\nu_j t)\exp\left[-\left(\frac{t - \epsilon_j}{\tau}\right)^2\right] }. \nonumber
\end{eqnarray}
This modulation leads to an $A_x\left( t \right) $ with Gaussian envelopes $G\left( t\right) = \exp\left[-\left(\frac{t - \epsilon_j}{\tau}\right)^2\right]$ peaking at $\epsilon_j$ within the pulse duration $\tau$.
We numerically solve Eqs.~(\ref{OBE21}-\ref{OBEprobeL}) with parameters 
$\beta_1 = \beta_2 = 0.055$,
$\nu_1 = 490$ rad$\cdot$kHz for $\Psi_2 \rightarrow \Psi_3$ transition, and
$\nu_2 = 684$ rad$\cdot$kHz for $\Psi_1 \rightarrow \Psi_3$ transition.
Fig.~\ref{fig3}(b) illustrates the dynamics of fidelity, and Fig.~\ref{fig3}(c) shows the optimized modulation envelope $G\left(t \right)$ for the need of adiabaticity \cite{vitanov2017, bergmann2019}  with $\epsilon_1 = 0.85$ ms,
$\epsilon_2 = 1.15$ ms, and
$\tau = 0.3$ ms.
The alternation of $F_1$ (red-solid line) and $F_2$ (green-dashed line) midway  occurs  through the two pulsed modulations, indicating that $\rho_{21}$, initially in mode $\Psi_1$, gradually becomes mode $\Psi_2$ under the action of the synthetic  $A_x$. 
Moreover, we illustrate the STIRAP process with snapshots of $\vert\rho_{21}\vert^2$ at different times: $t=0.1$ms in Fig.~\ref{fig3}(d), $t=1.03$ ms in (e), and $t=1.9$ ms in (f).
At the onset of the STIRAP, Fig.~\ref{fig3}(d) shows $\vert\rho_{21}\vert^2$ starting to deviate from its initial position at $\mathbf{x}=-5$mm, moving toward the right side of $U(\mathbf{x})$. 
This transition is accompanied by a decrease in $F_1$ 
and a corresponding increase in $F_2$. 
At $t=1.03$ ms, as shown in Fig.~\ref{fig3}(e), $\vert\rho_{21}\vert^2$ forms an even double-hump structure, indicating that the dark-state polarization is in a  superposition of $\Psi_1$ and $\Psi_2$. 
Ultimately, $\rho_{21}$ completely becomes $\Psi_2$ and rightward moves to $\mathbf{x}=5$mm as depicted by Fig.~\ref{fig3}(f). 
Our result highlights a potential application as a photonic-memory optimizer and modulator. For instance, in Fig.~\ref{fig3}(b-e),  photonic data initially is sent along y direction, stored at $\mathbf{x}=-5$mm, and then coherently reallocated to  $\mathbf{x}=5$ mm by transversely  applying $A_x$. 
Notably, our method preserves both the amplitude and phase information during the relocation process.
Although above STIRAP scheme   performs well, its double-pulse modulation for each control laser is complex. The time consuming of typical STIRAP is typically limited by the double-pulse sequence,  such as the 1 ms-long modulation shown in Fig.~\ref{fig3}(c).
An approach faster and simpler than the forward mentioned above double-pulse modulation can be considered by reallocating the pre-stored photonic data by the use of a single-pulse modulation for each control laser, expressed as:
\begin{eqnarray}
\Omega_c^R\left( t\right)  &=& \frac{\Omega_c}{\sqrt{2}} \sqrt{1 + \beta \sin(\nu t)\exp\left[-\left(\frac{t - \epsilon} {\tau}\right)^2\right] }, \nonumber\\
\Omega_c^L\left( t\right) &=& \frac{\Omega_c}{\sqrt{2}} \sqrt{1 - \beta \sin(\nu t)\exp\left[-\left(\frac{t - \epsilon} {\tau}\right)^2\right] }. \nonumber
\end{eqnarray}
Using parameters $\left(C_1,\alpha,\lambda_{f},\phi \right) = (6.12 \times 10^{-4} \frac{\Gamma}{\text{mm}^2},$
$\sqrt{50} \text{ mm}, 11.56\text{ mm}, 6.25)$,
we design a symmetric double-well potential, shown as the gray-solid line in Fig.~\ref{fig4}(a), with
eigen-angular frequencies
$\omega_1 = 456$ rad$\cdot$kHz, 
$\omega_2 = 481$ rad$\cdot$kHz, and 
$\omega_3 = 1222$ rad$\cdot$kHz.
This potential results in two nearly degenerate ground modes, $\Psi_1$ (red-solid line) and $\Psi_2$ (green-dashed line), with $\omega_1 \simeq \omega_2$, enabling a single-pulse modulation scheme.
The system maintains a $\Lambda$-type structure, as revealed by 
$\vert\Omega_{21}\vert = 18$ rad$\cdot$kHz, $\vert\Omega_{31}\vert = 378$ rad$\cdot$kHz, and $\vert\Omega_{32}\vert = 440$ rad$\cdot$kHz, ensuring inefficient direct coupling between the two ground modes.
We numerically solve Eqs.~(\ref{OBE21}-\ref{OBEprobeL}) with parameters 
$\beta = 0.212$,
$\nu = 1442$ rad$\cdot$kHz for both $\Psi_2 \rightarrow \Psi_3$ and
$\Psi_1 \rightarrow \Psi_3$ transition,
$\epsilon = 0.5$ ms,
and
$\tau = 0.15$ ms.
Fig.~\ref{fig4}(b-f) show that a single-pulsed modulation $A_x$ successfully induces coherent transfer between  two ground modes and reallocates the stored photonic data.
Notably,  in contrast to Fig.~\ref{fig3}(c),  the single-pulse $G\left( t\right) $ in Fig.~\ref{fig4}(c) spends only $0.15$ms to rightward shift the dark-state polarization from $\mathbf{x}=-5$mm to  $\mathbf{x}=5$mm.
Compared to the double-pulse scheme in Fig.~\ref{fig3}, this approach significantly simplifies the memory allocation procedure and reduces processing time by a factor of 6.7.

Our three-level-$\Lambda$-type scheme in Fig.~\ref{fig1}(a) can potentially be realized in a cold $^{133}$Cs atomic medium with its $D_1$  lines \cite{Hsiao2018}:
$\vert 1 \rangle = \vert 6S_{1/2}, F=3, m_F = 0 \rangle$, 
$\vert 2 \rangle = \vert 6S_{1/2}, F=4, m_F = -1 \ \mathrm{or} \ 1 \rangle$, and 
$\vert 3 \rangle = \vert 6P_{1/2}, F=3, m_F = 0 \rangle$.
One can engineer $ U(\mathbf{x})$  by Zeeman energy shift and integrating a periodic magnetic field \cite{yang2017} with the Ioffe-Pritchard trap \cite{Petrich1995}. A double-well potential had been realized by using a Ioffe-Pritchard trap with a time orbiting potential \cite{tiecke2003} and atom chip traps \cite{Esteve2005}.
The above levels $\vert 1 \rangle$ and $\vert 3 \rangle$ with $m_F = 0$ lead to a constant $\Delta_p$, and $\vert 2 \rangle$ with $m_F = \pm1$ allows for the x-dependent  $\Delta_c$.
All present results, with or without the inclusion of $\gamma$, show no significant differences, as the last diffusion term in \eqref{schrodinger}  contributes to dephasing at the same order of magnitude as $\gamma$.
The most probable atomic speed 0.19 m/s of the Maxwell-Boltzmann distribution at temperature 300 $\mu$K leads to Doppler shifts $\pm 0.04\Gamma$ on the $D_1$ lines. This causes $\mp 0.9\%$ variation in the effective mass for $\Delta_p = 4.6\Gamma$, and approximately $\pm 0.3\%$ variations in the transition frequency between DSMs.

In conclusion, we present a method to generate and coherently control DSMs using EIT. 
Our approach leverages the versatile EIT-based Schr\"odinger equation simulator proposed in \cite{Kuan2023}.
A key feature of this scheme is its ability to manipulate the DSMs through a combination of synthetic scalar and vector potentials. 
To demonstrate the potential of our scheme, we show the Rabi oscillations between two DSMs and STIRAP within a three-mode  $\Lambda$-type configuration.
This dynamic control enables the exploration of a wide range of quantum mechanical phenomena and supports practical applications, such as photonic-memory optimization and modulation. For instance, our method allows for the coherent reallocation of  pre-stored photonic data across a memory medium, maintaining both amplitude and phase information.

This work is supported by National Science and Technology Council of Taiwan (Grant No. 
111-2923-M-008-004-MY3,
112-2112-M-007-020-MY3,
\&
113-2628-M-008-006-MY3).
I-K.L. is supported by the Science and Technology Facilities Council grant ST/W001020/1.
G.J. is supported by the Lithuanian Council of Research (Grant No.  S-LLT-22-2).
T.K. is supported by the Ministry of Education and Science of the Republic of Latvia (Grant No. LV-LT-TW/2024/11 “Coherent Optical Control of Atomic Systems”).

\bibliography{LGEIT}

\end{document}